\documentclass[IOP]{emulateapj} \usepackage{amssymb}
\usepackage{lscape} \usepackage{longtable}

 \shorttitle{The Redshift-Dependent Ly$\alpha$
  fraction in LBG samples}
\begin{document}

\shortauthors{Stark et al.}

\title{Keck Spectroscopy of Faint $3<z<7$ Lyman Break Galaxies:- II. A
  High Fraction of Line Emitters at Redshift Six}


\author {Daniel P. Stark\altaffilmark{1},  Richard
  S. Ellis\altaffilmark{2}, Masami Ouchi\altaffilmark{3,4,5,6} }

\altaffiltext{1}{Kavli Institute of Cosmology \& Institute of
  Astronomy, University of Cambridge, Madingley  Road, Cambridge CB3
  0HA}   \altaffiltext{2}{Department of Astrophysics, California
  Institute of Technology, MS 105-24, Pasadena, CA 91125}
\altaffiltext{3}{Institute for Cosmic Ray Research, University of Tokyo, 
Kashiwa 277-8582, Japan}
\altaffiltext{4}{Institute for the Physics and Mathematics of the 
Universe (IPMU),University of Tokyo, Kashiwa 277-8568, Japan}
\altaffiltext{5}{Observatories of the Carnegie Institution of
  Washington, 813 Santa Barbara Street Pasadena, CA 91101 USA}
\altaffiltext{6}{Carnegie Fellow}

\begin{abstract} 
 
As Lyman $\alpha$ photons are scattered by neutral hydrogen, a change
with redshift in the Ly$\alpha$ equivalent width distribution of
distant galaxies offers a promising probe of the degree of ionization
in the intergalactic medium and hence when cosmic reionization ended.
This simple test is complicated by the fact that Ly$\alpha$ emission
can  also be affected by the evolving astrophysical details of the
host galaxies.  In the first paper in this series, we demonstrated
both a luminosity and redshift dependent trend in the fraction of
Ly$\alpha$ emitters seen within color-selected `Lyman-break'
galaxies (LBGs) over the range $3<z<6$; lower luminosity galaxies and
those at higher redshift show an increased likelihood of strong
emission. Here we present the results from much deeper 12.5 hour
exposures with the Keck DEIMOS spectrograph focused primarily on LBGs
at $z\simeq 6$ which enable us to confirm the redshift  dependence of
line emission more robustly and to higher redshift than was  hitherto
possible. We find $54\pm 11$\% of faint $z\simeq 6$ Lyman break
galaxies show  strong ($\rm{W_{Ly\alpha,0}}>25$~\AA) emission, an
increase of 1.6$\times$ from a similar sample observed at $z\simeq 4$.  With
a total sample of 74 $z\simeq 6$ LBGs, we determine the
luminosity-dependent Ly$\alpha$ equivalent width distribution.
Assuming continuity in these trends to the new population  of $z\simeq
7$ sources located with the Hubble WFC3/IR camera, we predict that
unless the neutral fraction rises in the intervening 200 Myr,  the
success rate for  spectroscopic confirmation using Ly$\alpha$ emission
should be high.

\end{abstract} 

\section{Introduction}
\label{sec:intro}

The reionization of neutral hydrogen in the intergalactic medium (IGM)
was a  landmark event in cosmic history, rendering the Universe
transparent to UV  photons and dramatically reducing the star
formation efficiency in dwarf galaxies.   In spite of its importance,
there are few robust constraints on when reionization occurred.
Polarization measures of the microwave background radiation
\citep{Larson10} demonstrate scattering by free electrons in the
redshift  range $7<z<20$ but do not describe the evolving neutral
fraction, $x_{HI}$.  Absorption line spectra of high-$z$ quasars are
largely sensitive to the very late stages ($x_{HI}\simeq 10^{-3}$) of reionization \citep{Fan06}, and progress has been slow
due to the paucity of sources so far detected beyond  $z\simeq 6.5$. 

One of the most promising probes of reionization with current
facilities is through the study of Ly$\alpha$ emission from star
forming galaxies.  Since Ly$\alpha$ photons are resonantly scattered
by neutral hydrogen, the abundance of Ly$\alpha$ emitters should
decrease as observations probe into the era where there are pockets of
neutral gas.  Studies of the redshift-dependent luminosity function
(LF) of Ly$\alpha$ emitters (LAEs) selected via narrowband filters
have revealed a possible decline in abundance between $z=5.7$ and
$z=7.0$ (\citealt{Kashikawa06,Iye06,Ota08,Ouchi10}), offering
tantalizing evidence that this short time interval ($\simeq$200 Myr)
may correspond to one during which there is some  evolution in the
neutral fraction.  But since a number of astrophysical factors can
also affect the presence of Ly$\alpha$ emission, it may be dangerous
to directly link  evolution in the Ly$\alpha$ LF to reionization
(e.g., \citealt{Dayal10}).  These factors include time-dependent
changes in the host galaxy number density, dust obscuration and
interstellar gas content and kinematic properties. By enlarging the
LAE samples, it may be possible  to bypass some of these complications
by testing for the expected change in their spatial clustering and
line profiles as the neutral era is  entered \citep{Ouchi10}.

A complementary approach introduced in \citet{Stark10a} (hereafter
Paper I) is to spectroscopically measure the fraction of strong
Ly$\alpha$ emitters  within the {\it color-selected Lyman Break Galaxy
  (LBG) population}.  By tracing the redshift-dependent fraction, the
host galaxy number density is not a factor. Evolution in dust
obscuration can be independently tracked using the continuum colors
and ISM kinematics through deep spectroscopy
\citep{Steidel10,Bouwens09a,Bouwens10a,Vanzella09}.  Although
demanding  observationally, high throughput spectrographs such as
FORS2 on the ESO Very Large Telescope and DEIMOS on the Keck II
telescope have enabled progress in recent years (Paper I,
\citealt{Vanzella09}).  With the additional information on the host
galaxies possible for the LBG population, we can hope to more reliably
link any redshift-dependence in the Ly$\alpha$  fraction to ionization
changes in the IGM. Most importantly of all, the proposed approach can
now be readily extended to $z\simeq 7$-8 and beyond given the
availability of LBG samples at these early epochs following the advent
of the WFC3/IR camera onboard Hubble Space Telescope (HST,
e.g. \citealt{Bouwens10a}).

In Paper I we introduced a large Keck spectroscopic survey of $z>3$
LBGs and demonstrated the practical details of the above method
through analyses of the Ly$\alpha$ fraction ($X_{Ly\alpha}$) in
B-band  ($z\simeq 4$) and V-band ($z\simeq 5$) dropouts to which we
added a sample  of $i'$-band ($z\simeq 6$) dropouts drawn  from other
programs (e.g., \citealt{Vanzella09},  Bunker et al. 2010, in prep).
Correcting for  minor magnitude and redshift-dependent biases in
completeness and contamination, we determined the luminosity and
redshift dependence of  $X_{Ly\alpha}$ over $3<z<6$. Since the IGM is
known to be highly ionized  over this interval, this dataset enabled
us to explore the importance of factors other than the IGM neutral
fraction. We found that galaxies with lower rest-frame UV continuum
luminosities exhibit Ly$\alpha$ emission  more frequently than
luminous systems.  Correlations between line strength and  UV
continuum slope suggest reduced dust obscuration is the primary cause.
The data also suggest an increase in $X_{Ly\alpha}$ with redshift
(dx$\rm{_{Ly\alpha}/dz \simeq 0.05 \pm 0.03}$), as originally claimed
based on the relative evolution of the UV luminosity function of  LAEs
and LBGs over this redshift range \citep{Ouchi08}.  However, since the
size of our archival $z\simeq 6$ samples were considerably smaller
than that of our lower redshift database, the constraints on the
redshift evolution were primarily derived from data spanning only the
300 Myr between $z\simeq 4$ and $z\simeq 5$.

Ideally one would construct the full equivalent width (EW)
distribution function of Ly$\alpha$ emission for a range of UV
luminosities at the highest redshift where the IGM is known to be
highly  ionized, i.e. $z\simeq 6$. This could then form the basis for
comparisons with spectroscopic data at $z>7$ where reionization may
be incomplete. With this motivation, we have thus extended the sample
introduced in Paper I, through ultra-deep spectroscopy of a sample of
$i'$-band dropouts in the GOODS North field. The increased sample size
and deep spectroscopic exposures provide statistically-significant
constraints on the  EW distribution of feeble sources, ensuring an
adequate basis for comparisons with higher redshift spectroscopic
samples.

Throughout the paper, we adopt a $\Lambda$-dominated, flat universe
with $\Omega_{\Lambda}=0.7$, $\Omega_{M}=0.3$ and
$\rm{H_{0}}=70\,\rm{h_{70}}~{\rm km\,s}^{-1}\,{\rm Mpc}^{-1}$. All
magnitudes in this paper are quoted in the AB system \citep{Oke83}.

\section{Observations}

\begin{figure}
\begin{center}
\includegraphics[width=0.47\textwidth]{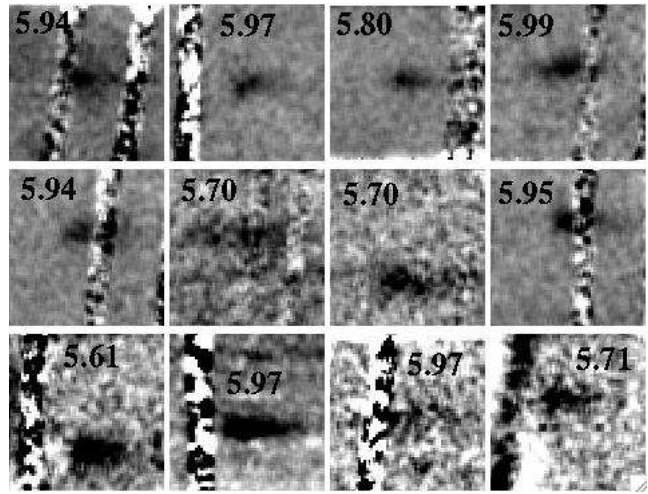}
\caption{Montage of 2-D Ly$\alpha$ detections in the April 2010 DEIMOS
  run targeting $i'$-band dropouts in the GOODS North field.  The
  color  scale is inverted with black corresponding to positive
  flux. Each cutout spans 7.1 arcsec $\times$ 28 ~\AA.  The $i'$-band
  dropouts that we observed are faint, with optical magnitudes
  spanning  $26.0 < z_{850}< 27.5$.   The Ly$\alpha$ detections are
  coincident with the spatial position of a UV continuum dropout
  satisfying the $i'$-band  dropout color criteria ($i'$-$z>1.3$ and
  containing no detections in deep B$_{435}$ or V$_{606}$
  imaging). Only  emission lines with $W_{Ly\alpha,0}>25$ \AA\ are
  included in the analysis in \S3 and \S4 (which excludes the second
  from right source in the bottom row), ensuring that the EW
  distribution  is derived from robust detections.  }
\label{fig:lya_idrops}
\end{center}
\end{figure}

Our dataset is primarily comprised of spectra obtained using the  DEep
Imaging Multi-Object Spectrograph (DEIMOS) at the  Nasmyth focus of
the 10 m Keck II telescope \citep{Faber03}. We direct the reader to
Paper I for a full description of our survey strategy.  In Paper I we
presented analysis of 513 DEIMOS spectra, including 268  unique
B-drops and 95 unique V-drops. To this we added publicly  available
spectra from the VLT/FORS2 survey of $z\simeq 4$, 5, and 6 
LBGs \citep{Vanzella09} and 2 unique $z\simeq 6$ LBGs from the Keck
survey of Bunker et al (2010, in prep).  The total 
sample drawn from Paper I is thus 351 B-drops, 151 V-drops, and  44
$i'$-drops.
 
The major step forward here is the inclusion of new $z\simeq 6$
spectra following ultra-deep Keck exposures of faint $i'$-band
dropouts in GOODS-North. The archival data in Paper I was mostly based
on the equivalent of 3-4 hour exposures with a 10 meter aperture. The
new sample consists of 23 $i'$-band dropouts with 12.5 hour exposures
(and an additional 7 with 3.67 hours of integration) enabling
constraints  to be placed further down the EW distribution at $z\simeq
6$ (allowing a  uniform sampling over the full redshift range) and
increasing the total $z\simeq 6$ LBG sample  by nearly $\simeq 70$\%
to 74 across both GOODS fields. 

\begin{figure*}
\begin{center}
\includegraphics[width=0.49\textwidth]{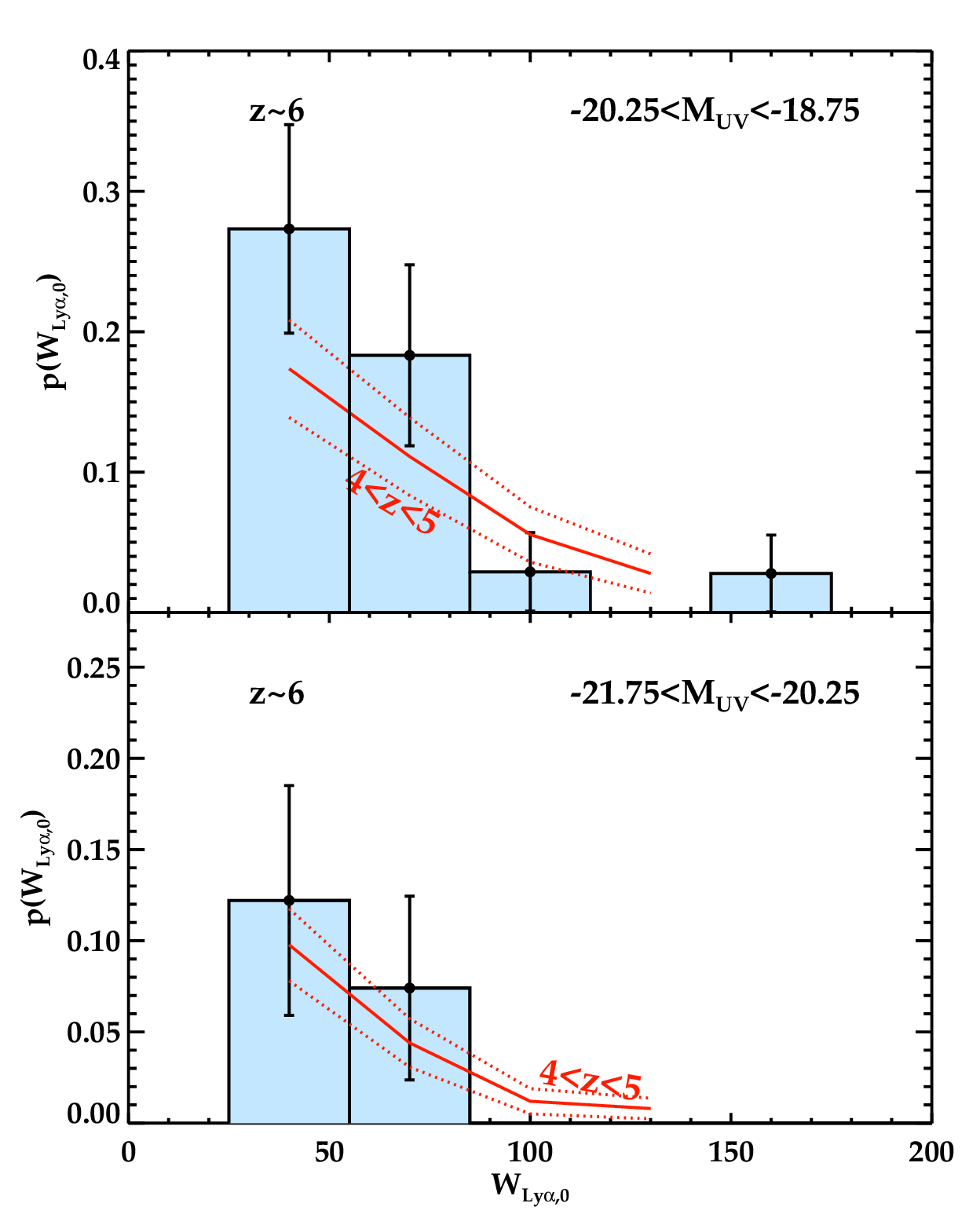}
\includegraphics[width=0.49\textwidth]{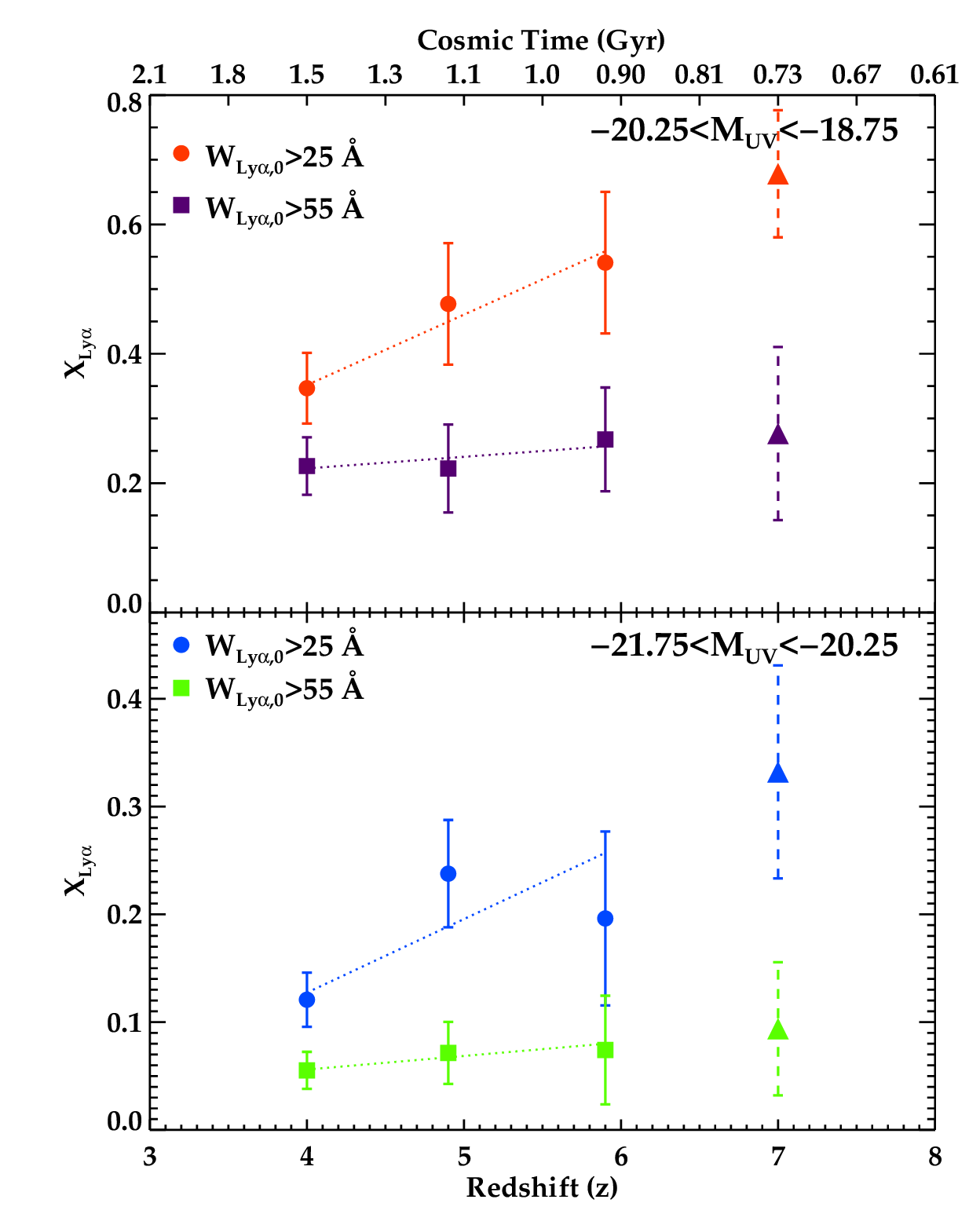}
\caption{ {\it Left:} The differential rest-frame Ly$\alpha$ EW
  distribution, $p(W_{Ly\alpha})$ (computed in bins spanning $\Delta
  W_{Ly\alpha,0}$=30~\AA) for star-forming galaxies at $z\simeq 6$ in
  two luminosity ranges ($-21.75<M_{UV}<-20.25$ on bottom and
  $-20.25<M_{UV}<-18.75$ on top).  Overplotted in red is the
  Ly$\alpha$ EW distribution for LBGs at $4<z<5$ derived from the
  sample in Paper I (dotted lines provide 1$\sigma$ uncertainties).
  {\it Right}:  Evolution in the overall fraction of Ly$\alpha$
  emitters ($X_{Ly\alpha}$)  in the LBG population over $4<z<6$.
  Luminous LBGs are considered in the bottom panel, and less luminous
  systems in the top panel. In each panel, we derive the Ly$\alpha$
  fraction of LBGs with Ly$\alpha$ EWs larger than 25~\AA~ (circles)
  and 55~\AA\ (squares). Assuming a linear relationship between
  X$_{Ly\alpha}$  and $z$, we extrapolate  these trends to $z\simeq 7$
  (triangles with dashed-line error bars).  }
\label{fig:lya_frac}
\end{center}
\end{figure*}

The new data were taken during April 2010. Over 11-12
April, we obtained 12.5 hours of on-source integration in good seeing
($<$0\farcs8) for one mask containing 23 $i'$-band dropouts.   Over
13-14 April, we obtained 3.67 hours of integration on a separate  mask
containing 7 $i'$-band dropouts.    For both masks, we used the 830
line mm$^{-1}$ grating, typically providing spectral coverage between
7000 ~\AA~ and 10400 ~\AA.  Slit lengths were generally $\simeq
7$\arcsec, and slit widths were 1\arcsec.  Skylines are measured to
have a Gaussian $\sigma$ of 1.1~\AA.  Reduction was performed using
the spec2d IDL pipeline developed for the DEEP2 survey
\citep{Davis03} \footnote{The spec2d pipeline can be downloaded at
  http://juno.as.arizona/cooper/deep/spec2d/}.  Wavelength calibration
was performed using Ne+Xe+Cd+Hg+Zn reference arc lamps.  As in Paper
I, we flux calibrate our data using the spectra of alignment stars
included  on the  slitmask (observed in 2\arcsec\ by 2\arcsec\ boxes).
We compared this calibration to that obtained using spectroscopic
standard stars and found it to be  consistent to within $\pm 20$\%
(with no  significant systematic offset) for the alignment stars.
Using the flux calibration, we computed our survey sensitivity as a
function of wavelength.  The 5$\sigma$ limiting line flux is 3.1$\pm
0.5 \times$10$^{-18}$ erg cm$^{-2}$ s$^{-1}$ (assuming a  range of
Ly$\alpha$ line widths typical of our LBG samples), implying that we
should be able to detect Ly$\alpha$ with rest-frame equivalent widths
of greater than $20\pm 3$ \AA\ for $i'$-drops with $z_{850}\simeq 27$.

\section{Analysis}

We searched for Ly$\alpha$ emission at the spatial position of the
targeted LBGs in the Keck spectra. Line fluxes and EWs were calculated
following the procedures discussed in Paper I.  We account for the
effects of line contamination and  Ly$\alpha$ forest absorption
(estimated using relations presented in \citealt{Meiksin06}) on the
observed  z$_{850}$-band fluxes.   Of the 23 $i'$-band dropouts for
which we obtained ultra-deep  spectra, 11 show Ly$\alpha$ emission,
while 2 of the 7 $i'$-drops for which we obtained 3.67 hour
integrations show Ly$\alpha$  (Figure \ref{fig:lya_idrops}).  These
results imply a large fraction of $i'$-drops have prominent Ly$\alpha$
emission.   The rest-frame EWs for the $i'$-drops range  between 9.4
~\AA~ and 350~\AA.  The vast majority of the emission lines are
detected with high significance. Even so,  we take a conservative cut,
limiting  our analysis to those sources with rest-frame EWs greater
than 25~\AA\ and $S/N>7$.   This excises the Ly$\alpha$ detection in
the middle right bottom panel of Figure \ref{fig:lya_idrops}.  As a
result, even among the faintest sources, the emission lines used in
our analysis are very  confidently detected ($<S/N>$=18), removing
concern regarding spurious features.    

As in Paper I, we determine the completeness of our Ly$\alpha$
detections as a function of  wavelength by adding and recovering fake
emission lines at random positions across the 2-D spectra. We compute
the Ly$\alpha$ recovery rate as a function of absolute magnitude and
wavelength for all masks observed (including those in Paper I) and
make appropriate corrections.  This test demonstrates that in our deep
12.5 hour mask we are $>$90\% complete to lines with
W$_{Ly\alpha,0}>50$ \AA~ even for the faintest $i'$-drops on our mask
($z_{850}\simeq 27$). For lines with W$_{Ly\alpha,0}\simeq 20$~\AA,
the completeness implied by our simulations is $\simeq 75-80$\% for
sources in the  faintest magnitude bin covered by our $z\simeq 6$
spectra.   The completeness is of course lower on the DEIMOS mask
observed for only 3.67 hours, reaching below $\simeq 50$\% for faint
sources with W$_{Ly\alpha,0}\simeq 20$~\AA, and we therefore  do not
include these sources when computing  the fraction of LBGs with low EW
Ly$\alpha$ emission.

An additional concern is that the color-cut and $z$-band selection of
$i'$-band dropouts are affected by Ly$\alpha$ emission and Ly$\alpha$
forest absorption.  We investigate the extent to which these effects
transform the observed EW distribution  using Monte Carlo simulations.
We create a large sample ($>$10$^6$) of artificial galaxies with
intrinsic absolute magnitudes (normalized  at 1500~\AA) spanning
$-21.5 < M_{UV} < -18.5$ and redshifts spanning $5.6<z<6.5$.  The
intrinsic luminosity distribution of the fake galaxies matches the
observed $i'$-drop luminosity function (e.g., \citealt{Bouwens06a}).
For the spectral shape, we use synthetic templates  (Charlot \&
Bruzual 2010, in preparation) with parameters fixed to those  which
provide reasonable fits for similarly bright $i'$-dropouts
(e.g. \citealt{Stark09}).  Changing  these parameters to other
reasonable  values does not affect our results.  We attach Ly$\alpha$
luminosities to each of the galaxies according to an assumed
Ly$\alpha$ EW distribution (which we describe below) and we also
account for Ly$\alpha$ forest absorption using the relations presented
in \citet{Meiksin06}.  Finally we derive  $i'_{775}$ and
$z_{850}$-band  magnitudes from the model SEDs and construct an
artificial sample  of $i'$-drops which satisfy the color criteria and
$z$-band magnitude  limit.  

We find that the output EW distribution matches the input EW
distribution  of galaxies at the mean redshift of the $i'$-drop
population.  For example,  if we adopt an input EW distribution with
the form $p(W_{Ly\alpha,0}$= exp[-$W_{Ly\alpha,0}$/$W_0$] and set
$W_0$=20.0~\AA, we find that the output  EW distribution is nearly
identical to the input distribution  ($W_0$=20.1~\AA).  It should be
noted that Poisson noise (which tends to scatter  faint sources toward
slightly brighter magnitudes) will alter the  EW distribution if the
intrinsic EW distribution is luminosity-dependent, as  suggested by
Paper I.  But this effect should occur at each redshift and  hence
should not affect the measured redshift evolution in the EW
distribution. 

\section{Results}

We now derive the EW distribution and Ly$\alpha$ fraction
($X_{Ly\alpha}$) for $z\simeq 6$ galaxies and compare with the lower
redshift samples of Paper I.  We group our $i'$-drop sample  into two
bins of rest-UV absolute magnitude, taking care to apply minor
corrections to the observed broadband magnitudes to compensate for the
effects of Ly$\alpha$ emission  and IGM Ly$\alpha$ forest absorption.
For galaxies without Ly$\alpha$ emission,  we correct for Ly$\alpha$
forest absorption statistically using the  redshift distribution
predicted from the Monte Carlo simulations in \S3. 

In Figure 2, we present our observed Ly$\alpha$ EW distribution, with
emission lines grouped in 30~\AA~ bins.  For the luminous sub-sample,
the distribution, p($W_{Ly\alpha,0}$), rises toward lower EW widths,
reaching p=$12\pm 6.8$\%  in the lowest EW bin considered. When
compared  to the EW distribution of luminous sources at $4<z<5$  (from
Paper I), we find that while Ly$\alpha$ is marginally more common in
each EW  bin at $z\simeq 6$, the uncertainties are too large to
distinguish the two distributions.  In contrast, in the lower
luminosity bin, the  EW distribution shows stronger positive evolution
from $4<z<5$, with Ly$\alpha$ considerably more prevalent among
$z\simeq 6$ LBGs. 

We next compute the LBG Ly$\alpha$ fraction by integrating the EW
distribution above 25~\AA~ and 55~\AA~ to yield the fractions
X$^{25}_{Ly\alpha}$ and X$^{55}_{Ly\alpha}$.  We group galaxies in the
same two luminosity bins as in the analysis above.  In the faint
subset, we find $54\pm 11$\% have W$_{Ly\alpha,0}>25$~\AA~ and $27\pm
8.0$\% have W$_{Ly\alpha,0}>55$~\AA.  Luminous galaxies exhibit
Ly$\alpha$ emission less frequently, with $20\pm 8.1$\% and $7.4\pm
5.0$\% observed with Ly$\alpha$ emission in excess of 25 and 55~\AA.
Combining our results with those from Paper I, we find that the
fraction of Ly$\alpha$ emitters among the LBG population  increases
with redshift for lower luminosity galaxies.  Assuming a linear
relationship between X$_{Ly\alpha}$ and redshift, we find
dX$^{25}_{Ly\alpha}$/dz = $0.11 \pm 0.04$.  In contrast, less redshift
evolution is seen in the larger EW bin (dX$^{55}_{Ly\alpha}$/dz =
$0.018 \pm 0.036$),  consistent with the findings from Paper I.
Similar (albeit  noisier) trends are seen in the more luminous
sub-sample, with  the lowest EW bin showing the strongest indications
of positive evolution  with redshift.

This improved determination of the EW distribution and Ly$\alpha$
fraction for LBGs at $z\simeq 6$ is a necessary step toward providing
the essential baseline for predicting the outcome of spectroscopic
campaigns beyond $z\simeq 7$ and interpreting any downturn in the
Ly$\alpha$  fraction that may be  associated with reionization (see \S
5).

\begin{figure}
\begin{center}
\includegraphics[width=0.49\textwidth]{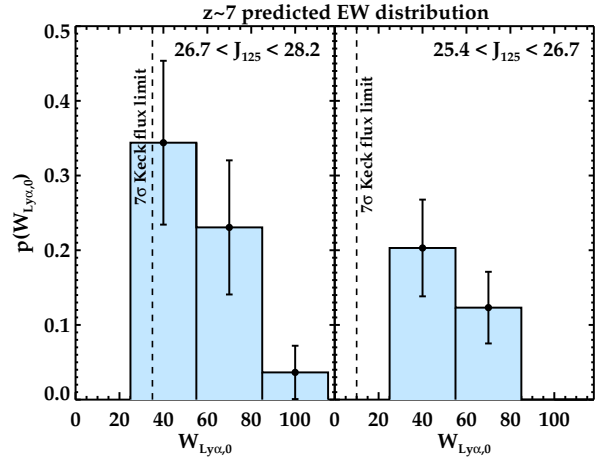}
\caption{Predicted rest-frame Ly$\alpha$ EW distribution (in bins of
  $\Delta$W$_{Ly\alpha,0}$=30~\AA) for $z\simeq 7$ LBGs based on an
  extrapolation of trends from Fig.~ 2 assuming that the Ly$\alpha$
  fraction increases linearly with redshift.  Uncertainties are based 
  on statistical error in our lower redshift samples. The dashed line
  indicates the limits that could be reached $\simeq 4$  hours of
  integration with Keck/NIRSPEC.  Significant deviations below this
  prediction  may arise if the IGM is partially neutral.  }
\label{fig:lya_z7}
\end{center}
\end{figure}

\section{The Expected Visibility of Ly$\alpha$ Emission in $z>7$ LBGs}

Our new results, taken together with those in Paper I, now suggest
that $\simeq 54$\% of moderately faint ($-20.25 < M_{UV}< -18.75$)
$z\simeq 6$ LBGs exhibit very strong Ly$\alpha$ emission.  In Paper I,
we argued that both the redshift and luminosity dependence of the
Ly$\alpha$ fraction was likely due in large part to variations in dust
obscuration as evidenced by the correlation between Ly$\alpha$ EW and
the rest-frame UV slope, $\beta$. Recent analyses of the colors of the
$z\simeq 7$ LBGs indicate that these systems are yet bluer than those
at $z\simeq 6$ \citep{Bouwens10a}, implying even less or no dust
obscuration.  Hence it seems likely that the redshift trend in the
Ly$\alpha$ fraction presented in Figure \ref{fig:lya_frac} should
continue to $z\simeq 7$ and that Ly$\alpha$ should be
readily detectable in sufficiently deep spectroscopic  campaigns.

Given the short cosmic  time spanning $6<z<7$ ($\simeq 170$ Myr), it
seems plausible to use the EW distribution and Ly$\alpha$ fractions
presented in Figure~\ref{fig:lya_frac} to  predict the expected
Ly$\alpha$ visibility for  sources at $z\simeq 7$, assuming Ly$\alpha$
flux is not significantly  attenuated by neutral hydrogen in the IGM.
Motivated  by the blue $z\simeq 7$ UV slopes discussed above, we
extrapolate the  evolution in X$_{Ly\alpha}$ to
$z\simeq 7$ (Figure 2).  For low luminosity sources, this results in a
small increase in the Ly$\alpha$ fraction  ($\Delta
X^{25}_{Ly\alpha}=0.14$) which we divide into the three EW bins using
weights set by $p(W_{Ly\alpha,0})$.  We  follow the same procedure for
the luminous sources.    The results, presented in Figure 3, suggest a
survey of $\simeq 20$-30 galaxies drawn from the now-available WFC3/IR
target list  (e.g., \citealt{McLure10,Bouwens10d,Bunker10}) would
yield interesting results.  While uncertainty in the observed
Ly$\alpha$ trends and their  extrapolation to $z\simeq 7$ obviously
affects our prediction, it seems  clear that Ly$\alpha$ should be
common in $z\simeq 7$ samples if  the IGM is highly ionized.
The failure to detect emission in such a sample might therefore be a
strong indicator of a rising neutral fraction beyond $z\simeq 6$ as
claimed originally by \citet{Kashikawa06} from the luminosity function
of LAEs.
 
How practical is a search for line emission to the EW limit of 20~\AA~
discussed  above? In terms of an integrated line flux $F_{Ly\alpha}$,
the EW limit corresponds to $F_{Ly\alpha} \simeq 3(7)
\times$10$^{-18}$ erg cm$^{-2}$ s$^{-1}$ for $z\simeq 7$  galaxies
with M$_{UV}$=-20(-21) (corresponding to galaxies with apparent AB
magnitudes of $J \simeq 27$ and 26,  respectively).  Such line flux
limits are feasible with spectrographs on 8-10 meter telescopes  such
as the Near InfraRed SPECtrograph (NIRSPEC) on Keck II
\citep{McLean98}.  Earlier work with NIRSPEC has reached such a limit
at 5-$\sigma$ significance between the atmospheric  sky lines in the
$Y$ and $J$-band in $\simeq 4$ hours 
\citep{Stark07b,Richard08}.   As more  multi-object infrared
spectrographs become available, it will be feasible to observe many
$z>7$  sources simultaneously, allowing ultra-deep exposures of galaxies at least as faint as $M_{UV}\simeq $~-19.  

\section{Conclusions}

We present new ultra-deep spectroscopic observations of 30 $i'$-band
dropouts in GOODS-N using DEIMOS on the Keck II telescope.  By adding
these spectra to the large database of DEIMOS and FORS2 spectra of  B,
V, and $i'$-band dropouts discussed in Paper I, we demonstrate more
robustly the rise with redshift over $4<z<6$ in the fraction of  low
luminosity LBGs that show prominent Ly$\alpha$ emission. We also
derive a much-improved EW distribution of Ly$\alpha$ at $z\simeq 6$,
the highest redshift where  the intergalactic medium is known to be
highly ionized. Motivated by the continued blueward evolution of the
continuum UV slopes to $z\simeq 7$, we extrapolate the 
Ly$\alpha$  fraction trends presented in this paper to redshift 7, and
predict the likely success rate of recovering Ly$\alpha$ emission in
the new population of $z>7$ sources located with HST. Our
results suggest line emission should be readily detected given
adequate observing time and that quantitative results would therefore
test for the presence of neutral gas associated with the end of cosmic
reionization near $z\simeq 7$.

\subsection*{ACKNOWLEDGMENTS}
DPS acknowledges financial support from a postdoctoral fellowship from
the Science Technology and Research Council and a Schlumberger
Interdisciplinary  Research Fellowship at Darwin College, University of 
Cambridge. MO is grateful to financial support from Carnegie Observatories 
via the Carnegie fellowship.


\end{document}